\documentclass[twocolumn]{aastex63}

\usepackage{hyperref}
\usepackage{float}

\newcommand{\Kepler}{\textit{Kepler}}
\newcommand{\Spitzer}{\textit{Spitzer}}
\newcommand{\kt}{\textit{K2}}

\shorttitle{K2-138~g}
\shortauthors{Hardegree-Ullman et al.}

\begin{document}

\title{K2-138~g: \Spitzer\ Spots a Sixth Planet for the Citizen Science System}

\correspondingauthor{Kevin K. Hardegree-Ullman}
\email{kevinkhu@caltech.edu}

\author[0000-0003-3702-0382]{Kevin K. Hardegree-Ullman}
\affil{Caltech/IPAC-NExScI, M/S 100-22, 1200 E. California Blvd, Pasadena, CA 91125, USA}

\author[0000-0002-8035-4778]{Jessie L. Christiansen}
\affil{Caltech/IPAC-NExScI, M/S 100-22, 1200 E. California Blvd, Pasadena, CA 91125, USA}

\author[0000-0002-5741-3047]{David R. Ciardi}
\affil{Caltech/IPAC-NExScI, M/S 100-22, 1200 E. California Blvd, Pasadena, CA 91125, USA}

\author[0000-0002-1835-1891]{Ian J. M. Crossfield}
\affiliation{Department of Physics and Astronomy, University of Kansas, Lawrence, KS 66045}

\author[0000-0001-8189-0233]{Courtney D. Dressing}
\affiliation{Department of Astronomy, University of California, Berkeley, Berkeley, CA 94720}

\author[0000-0002-4881-3620]{John~H.~Livingston}
\affiliation{Department of Astronomy, University of Tokyo, 7-3-1 Hongo, Bunkyo-ku, Tokyo 113-0033, Japan}

\author[0000-0001-8736-236X]{Kathryn Volk}
\affil{Lunar and Planetary Laboratory, The University of Arizona, 1629 E University Blvd, Tucson, AZ 85721}

\author[0000-0002-0802-9145]{Eric Agol}
\affiliation{Astronomy Department and Virtual Planetary Laboratory, University of Washington, Seattle, WA 98195 USA}

\author[0000-0001-7139-2724]{Thomas Barclay}
\affiliation{NASA Goddard Space Flight Center, 8800 Greenbelt Rd, Greenbelt, MD 20771, USA}
\affiliation{University of Maryland, Baltimore County, 1000 Hilltop Cir, Baltimore, MD 21250, USA}

\author[0000-0002-3306-3484]{Geert Barentsen}
\affil{Bay Area Environmental Research Institute, P.O. Box 25, Moffett Field, CA 94035, USA}

\author[0000-0001-5578-1498]{Bj\"{o}rn Benneke} 
\affil{Department of Physics and Institute for Research on Exoplanets, Universit\'{e} de Montr\'{e}al, Montreal, QC, Canada}

\author[0000-0002-8990-2101]{Varoujan Gorjian}
\affiliation{Jet Propulsion Laboratory,
California Institute of Technology, 4800 Oak Grove Drive, Pasadena, CA 91109, USA}

\author[0000-0002-2607-138X]{Martti H. Kristiansen}
\affil{Brorfelde Observatory, Observator Gyldenkernes Vej 7, DK-4340 T\o{}ll\o{}se, Denmark}
\affil{DTU Space, National Space Institute, Technical University of Denmark, Elektrovej 327, DK-2800 Lyngby, Denmark}

\begin{abstract}
\kt\ greatly extended \Kepler's ability to find new planets, but it was typically limited to identifying transiting planets with orbital periods below 40 days. While analyzing \kt\ data through the Exoplanet Explorers project, citizen scientists helped discover one super-Earth and four sub-Neptune sized planets in the relatively bright ($V=12.21$, $K=10.3$) K2-138 system, all which orbit near 3:2 mean motion resonances. The \kt\ light curve showed two additional transit events consistent with a sixth planet. Using \Spitzer\ photometry, we validate the sixth planet's orbital period of $41.966\pm0.006$ days and measure a radius of $3.44^{+0.32}_{-0.31}\,R_{\oplus}$, solidifying K2-138 as the \kt\ system with the most currently known planets. There is a sizeable gap between the outer two planets, since the fifth planet in the system, K2-138~f, orbits at 12.76 days. We explore the possibility of additional non-transiting planets in the gap between f and g. Due to the relative brightness of the K2-138 host star, and the near resonance of the inner planets, K2-138~could be a key benchmark system for both radial velocity and transit timing variation mass measurements, and indeed radial velocity masses for the inner four planets have already been obtained. With its five sub-Neptunes and one super-Earth, the K2-138 system provides a unique test bed for comparative atmospheric studies of warm to temperate planets of similar size, dynamical studies of near resonant planets, and models of planet formation and migration.
\end{abstract}



\section{Introduction}
\label{sec:introduction}
The NASA \kt\ mission searched for exoplanets in different fields spanning the ecliptic plane, subsequent to the loss of two reaction wheels, which inhibited the \Kepler\ spacecraft's ability to precisely point at the original \Kepler\ field for extended durations. Using solar pressure and thrusters, \Kepler\ was able to point to fields along the ecliptic plane for a period of $\sim$83 days each before the spacecraft was rotated to prevent sunlight from entering the telescope \citep{Putnam2014, Howell2014}. \kt\ has so far enabled the discovery of 425 new planets and an additional 889 planet candidates.\footnote{\url{http://exoplanetarchive.ipac.caltech.edu/docs/counts_detail.html}, as of February 2021.}

K2-138 was the first \kt\ planet system discovered by citizen scientists through the Exoplanet Explorers\footnote{\url{https://www.zooniverse.org/projects/ianc2/exoplanet-explorers}} program on the Zooniverse\footnote{\url{https://www.zooniverse.org/}} \citep{Christiansen2018}. The citizen scientists were able to identify four sub-Neptune sized planets by visual inspection of the light curve. A closer inspection of the diagnostic plots from the \texttt{TERRA} algorithm\footnote{\url{https://github.com/petigura/terra}} \citep{Petigura2013a,Petigura2013b} elucidated a super-Earth interior to the orbits of the other four planets. Using \texttt{LcTools} \citep{Kipping2015,Schmitt2019}, \citet{Christiansen2018} also identified two additional transits 41.97 days apart, indicating a possible sixth planet for the system.

\citet{Lopez2019} obtained radial velocity (RV) measurements of K2-138 with HARPS, yielding mass measurements of $3.1\pm1.1$, $6.3^{+1.1}_{-1.2}$, $7.9^{+1.4}_{-1.3}$, and $13.0\pm2.0\,M_{\oplus}$ for planets b, c, d, and e, respectively. Precise masses for K2-138~f and the putative planet K2-138~g were not measured. K2-138~f has an orbital period of 12.76 days, about half of the $24.7\pm2.2$ day stellar rotation period, and its signal was likely absorbed by the Gaussian process regression used to remove stellar activity. This process also likely muted the signal of K2-138~g. \citet{Lopez2019} placed upper limits at 99\% confidence of 8.7 and 25.5 $M_{\oplus}$ on K2-138~f and g, respectively. Due to the near 3:2 orbital resonances, K2-138 is amenable to transit timing variation (TTV) measurements to constrain planet masses. Using their measured masses and assuming zero eccentricity, \citet{Lopez2019} computed TTV amplitudes between 2.0 and 7.3 minutes for the inner five planets, similar to the amplitudes computed by \citet{Christiansen2018}. Though \citet{Christiansen2018} were not able to detect significant TTVs in the 30 minute cadence \kt\ data, higher cadence observations with instruments such as CHEOPS, which were scheduled for late 2020 (Program ID 017 (EP); PI: T. Lopez), should allow TTV mass measurements of planets c, d, and e, making K2-138 an important benchmark system for comparing TTV and RV masses. Since RV mass measurements are currently limited to host stars brighter than $V\lesssim13$, TTVs enable mass measurements for a much wider pool of planets \citep{Holczer2016}. However, fewer than 10 systems have both RV and TTV mass measurements, and detection sensitivity may bias RV measurements for planets with orbital periods larger than 11 days \citep{Mills2017,Petigura2018}. These reasons highlight the importance of adding new TTV/RV benchmark systems in order to cross-check masses between measurement techniques.

In this paper we verify the outermost planet K2-138~g with an orbital period of $41.96645^{+0.00603}_{-0.00665}$ days. This adds to the nine systems with six or more planets currently known, makes K2-138 the \kt\ discovered system with the most planets,\footnote{\citet{Kruse2019} identified six candidate planets in EPIC 210965800, five of which have yet to be confirmed.} and yields one of the longest period \kt\ planets. Using the \textit{Spitzer Space Telescope}, we observed a third transit of K2-138~g within one hour of the time predicted from the \kt\ ephemeris. We present our observations and data reduction in Section~\ref{sec:obsdata} and discuss our results in Section~\ref{sec:discussion}.

\section{Observations and Data Reduction}
\label{sec:obsdata}

\subsection{Stellar Classification}
\label{sec:stellar}

\begin{figure}[ht]
    \centering
    \includegraphics[width=0.46\textwidth]{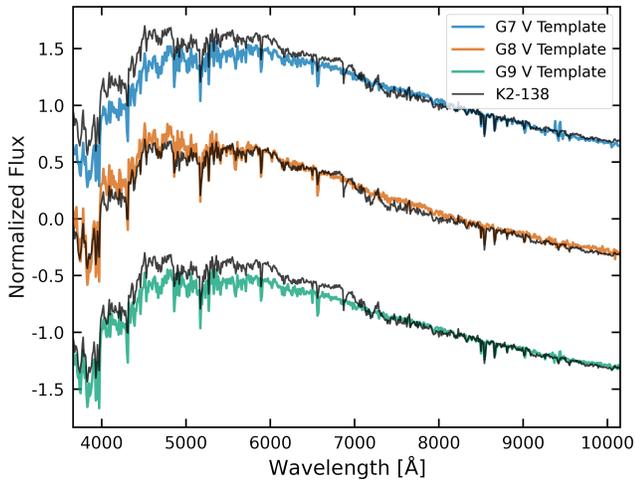}
    \caption{K2-138 spectrum (black) compared to G7 V (top), G8 V (middle), and G9 V (bottom) template spectra from \citet{Kesseli2017}. The G8 V template spectrum is the most similar to K2-138.} \label{fig:k2spec}  
		\vspace{-0.5em}
\end{figure}

We obtained a 0.38 to 0.7\,$\mu$m spectrum of K2-138 using the Goodman spectrograph \citep{Clemens2004} on the Southern Astrophysical Research Telescope (Program ID 2019A-0364; PI: K. Hardegree-Ullman), and a 0.7 to 2.4\,$\mu$m spectrum using the SpeX spectrograph \citep{Rayner2003} on the NASA Infrared Telescope Facility (Program ID 2017A-106; PI: K. Hardegree-Ullman). We followed the procedures outlined in \S\ 2.1-2.3 of \citet{Hardegree-Ullman2019} for observing the target and reducing the data. We compared the combined optical and infrared spectra between 0.38 and 1\,$\mu$m to optical SDSS spectral templates from \citet{Kesseli2017} following the procedures outlined in \S\ 4.1 in \citet{Hardegree-Ullman2020}, which yielded a spectral type of G8 V, consistent with the spectral type found by \citet{Lopez2019}. Figure~\ref{fig:k2spec} shows our 0.38 to 1\,$\mu$m spectrum compared to G7 V, G8 V, and G9 V template spectra.

\begin{deluxetable*}{cccccccc}
\tablecaption{K2-138 stellar parameters.\label{tab:starpars}}

\tablehead{\colhead{Spectral Type} & \colhead{$T_{\mathrm{eff}}$} & \colhead{$\log(g)$} & \colhead{[Fe/H]} & \colhead{$R_{\star}$} & \colhead{$M_{\star}$} & \colhead{Distance} & \colhead{Reference} \\
 & \colhead{K} & \colhead{$\log(\mathrm{cm\ s}^{-2}$)} & \colhead{dex} & \colhead{$R_{\odot}$} & \colhead{$M_{\odot}$} & \colhead{pc} & \colhead{}} 

\startdata
\nodata & $5189\pm156$ & $4.557\pm0.030$ & $0.01\pm0.12$ & $0.807\pm0.046$ & $0.870\pm0.052$ & $182.50\pm17.15$ & 1 \\
\nodata & $5228\pm76$ & $4.34\pm0.12$ & $0.26\pm0.20$ & \nodata & \nodata & $242.905\pm94.869$ & 2 \\
K1\,V $\pm$ 1 & $5378\pm60$ & $4.59\pm0.07$ & $0.16\pm0.04$ & $0.86\pm0.08$ & $0.93\pm0.06$ & $183\pm17$ & 3 \\
\nodata & $5110^{+209}_{-53}$ & \nodata & \nodata & $0.917^{+0.019}_{-0.070}$ & \nodata & $202.585^{+2.028}_{-1.989}$ & 4 \\
\nodata & $5281\pm129$ & $4.53\pm0.08$ & \nodata & $0.858\pm0.049$ & $0.91\pm0.11$ & $202.585\pm2.009$ & 5 \\
G8 & $5350\pm80$ & $4.52\pm0.15$ & $0.14\pm0.10$ & $0.86^{+0.03}_{-0.02}$ & $0.93\pm0.02$ & $201.66\pm6.38$ & 6 \\
G7 & $5303\pm138$ & $4.547\pm0.150$ & $0.035\pm0.235$ & $0.841^{+0.056}_{-0.050}$ & $0.911^{+0.412}_{-0.280}$ & $202.585^{+2.028}_{-1.989}$ & 7 \\
\hline
G8\,V $\pm$ 1 & $5283^{+122}_{-163}$ & $4.538^{+0.090}_{-0.159}$ & $0.134^{+0.128}_{-0.176}$ & $0.839^{+0.054}_{-0.051}$ & $0.916^{+0.078}_{-0.096}$ & $202.585^{+2.028}_{-1.989}$ & 8 \\
\enddata
\tablerefs{(1) EPIC \citep{Huber2016}, (2) RAVE DR5 \citep{Kunder2017}, (3) \citet{Christiansen2018}, (4) \textit{Gaia} DR2 \citep{Gaia2018,Bailer-Jones2018}, (5) TIC CTL v8.01 \citep{Stassun2019}, (6) \citet{Lopez2019}, (7) \citet{Hardegree-Ullman2020}, (8) This work.}
\vspace{-1.5em}
\end{deluxetable*}

In Table~\ref{tab:starpars} we list stellar parameters for K2-138~compiled from the Ecliptic Plane Input Catalog \citep[EPIC;][]{Huber2016}, RAVE DR5 \citep{Kunder2017}, \citet{Christiansen2018}, \textit{Gaia} DR2 \citep{Gaia2018,Bailer-Jones2018}, the \textit{TESS} Input Catalog Candidate Target List \citep[TIC CTL;][]{Stassun2019}, \citet{Lopez2019}, and \citet{Hardegree-Ullman2020}. The measured parameters are all consistent to within $1\sigma$, except for one measurement of $\log(g)$ which is within $2\sigma$. It is reassuring that different data sets and pipelines yield similar results, but when it comes to calculating planet parameters, small differences in stellar parameters can have a large impact. For large surveys and population studies of exoplanets, it is crucial to have a uniformly derived set of stellar parameters \citep[e.g.,][]{Fulton2017,Berger2020,Hardegree-Ullman2020}. For individual systems, however, it is typical to choose a single set of stellar parameters, which may be susceptible to systematic bias. Rather than cherry picking measurements from different references, we combined all the available measurements for $T_\mathrm{eff}$, $\log(g)$, [Fe/H], and $M_{\star}$. Instead of using a weighted mean, which would produce uncharacteristically small uncertainties,\footnote{A weighted mean ($\hat{\mu}=(\Sigma x_i/\sigma^2_i)/(\Sigma 1/\sigma^2_i)$, $\sigma^2(\hat{\mu}) = 1/(\Sigma 1/\sigma^2_i)$) would yield $T_\mathrm{eff} = 5300\pm35$ K, $\log(g) = 4.55\pm0.03$, [Fe/H] = $0.15\pm0.03$, and $M_{\star} = 0.92\pm0.02$. The uncertainties on the weighted mean parameters are $\sim$3-7 times smaller than the average individual measurement uncertainties. Our Monte Carlo uncertainties are much more conservative, and we believe they more accurately reflect typical measurement uncertainties for these parameters.} we instead employed the following Monte Carlo method. For each measurement with symmetric uncertainties, we randomly drew $10^4$ values from a Gaussian distribution, and for asymmetric uncertainties we drew $10^4$ values from a split normal distribution. The posterior distributions were concatenated and we took the median, 16th, and 84th percentiles of the resultant distribution as our measurement and errors.

We computed a bolometric magnitude ($M_{\mathrm{bol}}$) using the \textit{Gaia} distance of $202.585^{+2.028}_{-1.989}$\,pc reported by \citet{Bailer-Jones2018}, accounting for interstellar reddening with the \texttt{dustmaps} package \citep{Green2018}, and applying a bolometric correction found using \texttt{isoclassify} \citep{Huber2017}. We computed the bolometric luminosity using $L_{\mathrm{bol}}=L_0\times10^{-0.4M_{\mathrm{bol}}}$, where $L_0=3.0128\times10^{28}$\,W is the zero point radiative luminosity \citep{Mamajek2015}. From the Stefan-Boltzmann law we derived a new stellar radius of $0.839^{+0.060}_{-0.055}\,R_{\odot}$ with our Monte Carlo averaged effective temperature and bolometric luminosity. Stellar parameters for K2-138 are listed in Table~\ref{tab:starpars}.

\subsection{K2 Photometry}
\label{sec:k2}
EPIC 245950175 (K2-138) was observed with \kt\ in 30 minute long cadence mode during Campaign 12 between 2016 December 15 and 2017 March 04, with a five day gap in the data about two-thirds of the way through the campaign due to a spacecraft safe-mode event. K2-138 was included in Campaign 12 Guest Observing Programs 12049, 12071, 12083, and 12122 (PIs E. Quintana, D. Charbonneau, A. Jensen, and A. Howard). 

\begin{figure*}
    \centering
    \includegraphics[width=0.98\textwidth]{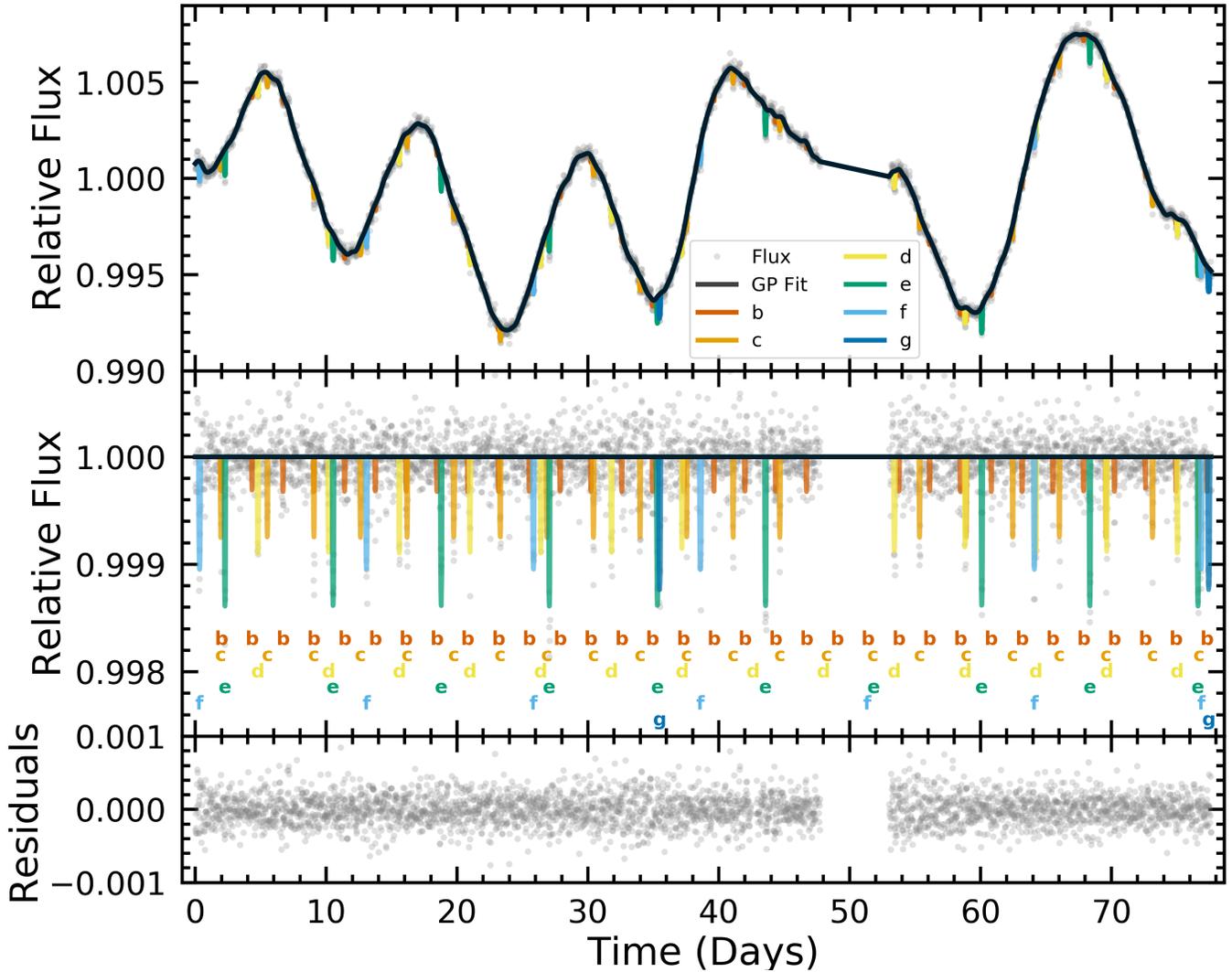}
    \caption{(Upper) The raw \kt\ light curve from the \texttt{k2phot} pipeline (gray points) with the Gaussian process fit (black) and six planet fits (colors). (Middle) The Gaussian process flattened light curve with planet fits. (Lower) Residuals after removal of all planet transit signals.} \label{fig:k2lc}
		\vspace{-0.5em}
\end{figure*}

For our analysis, we used the instrumental systematics-corrected light curve (Figure~\ref{fig:k2lc}) produced by the \texttt{k2phot\footnote{\url{https://github.com/petigura/k2phot}}} pipeline \citep{Petigura2015,Aigrain2016}. We compared the \texttt{k2phot} light curve to those produced by \texttt{EVEREST} \citep{Luger2016} and \texttt{K2SFF} \citep{Vanderburg2014}, and found that the \texttt{k2phot} light curves had the lowest overall RMS scatter and the fewest outliers. We first masked out data that was flagged in the \texttt{k2phot} pipeline as a thruster fire event or an outlier in background flux. Periodic transit signals were initially found by flattening the light curve with a Savitsky-Golay filter over a window of 101 points ($\sim$50 hours), then running a box least squares periodogram, iteratively masking out the higher signal-to-noise transits until there were no more convincing planet signals in the data. This search gave estimates of planet periods, transit times, and transit depths. Next, we made use of the \texttt{exoplanet\footnote{\url{https://exoplanet.dfm.io/en/stable/}}} toolkit to model stellar variability using a Gaussian process with a simple harmonic oscillator kernel, while simultaneously fitting planet transits as described in \citet{exoplanet:foremanmackey17}. In order to simultaneously fit the \kt\ and \Spitzer\ data (\S~\ref{sec:transit}), we used the flattened the light curve by subtracting the Gaussian process stellar model without fitting out the planet transits, which is shown in the middle panel of Figure~\ref{fig:k2lc}.

\begin{figure*}
    \centering
    \includegraphics[width=0.98\textwidth]{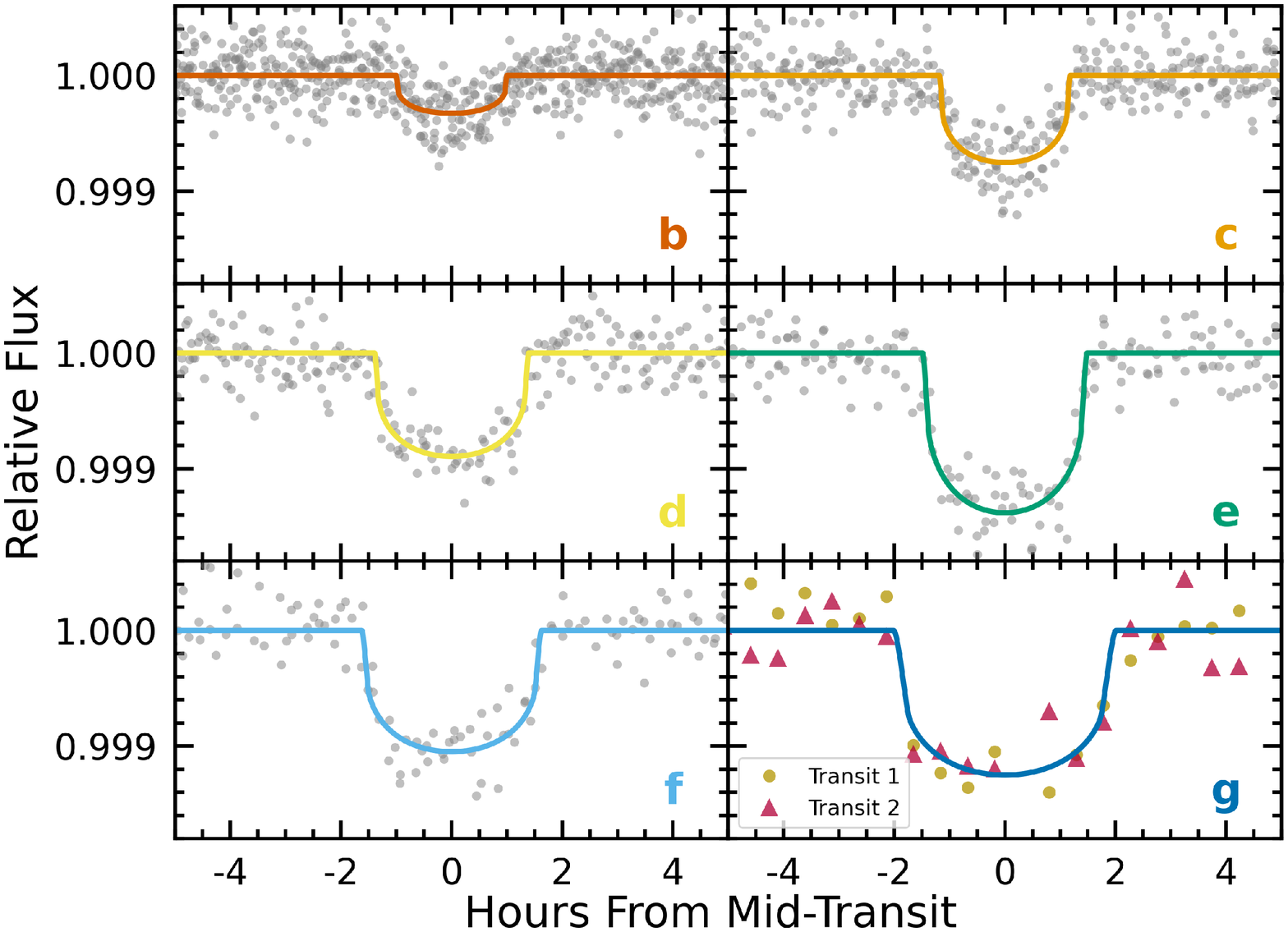}
    \caption{Phase folded \kt\ light curves for the six transiting K2-138 planets with best-fit transit models overlaid. The two transits for K2-138~g are shown in different colors and shapes in the lower right panel.} \label{fig:k2lc-phase}  
		\vspace{-0.5em}
\end{figure*}

\subsection{IRAC Photometry}
\label{sec:irac}
We observed K2-138 with the Infrared Array Camera (IRAC) on the \textit{Spitzer Space Telescope} (DDT 13253; PI J. Christiansen) at the predicted transit time of the putative sixth planet. We used Channel 2 (4.5\,$\mu$m) since it is less affected by intrapixel sensitivity variations than Channel 1 (3.6\,$\mu$m). The observation began with a 30 minute pre-observation stare which was discarded in the analysis, but included in the Astronomical Observation Request to allow the telescope and instruments to settle after slewing \citep{Grillmair2012}. To minimize the pixel-phase effect and achieve a pointing accuracy to within $\sim$0.1 pixel, we conducted pre-observations in peak-up mode using the Pointing Calibration and Reference Sensor \citep{Ingalls2012}.

Observations of K2-138 were conducted between 2018 March 15 and 2018 March 16 for a total duration of 11 hours centered near the predicted time of mid-transit from the \kt\ ephemeris of the sixth planetary signal found by \citet{Christiansen2018}. Individual frame exposure times were set to two seconds to stay in the linear regime of the detector for this bright target. The subarray mode was used to minimize readout times and data volume. In total, 19,840 individual frames were taken.

We performed centroiding and aperture photometry using \texttt{photutils} \citep{Bradley2019}, fitting a 2D Gaussian to each image. To select the optimal aperture radius, we computed photometry from the centroid positions using fixed radii between 1.5 and 3.0 pixels in 0.1 pixel increments. Background levels were found by the method described by \citet{Knutson2011}. This process entails masking out the regions within a radius of 12 pixels from the centroid along with the central two rows and columns, then finding the median background value of the pixels after clipping 3$\sigma$ outliers.

We modeled systematics in the \Spitzer\ light curves using pixel-level decorrelation \citep[PLD;][]{Deming2015}. PLD has become a premier technique for correcting \Spitzer\ systematics in planet transit analyses \citep[e.g.,][]{Beichman2016,Benneke2017,Dressing2018,Feinstein2019,Livingston2019,Berardo2019}, and was developed to account for intra-pixel sensitivity variations which produce intensity fluctuations in the photometry. In our analysis, we used PLD to model the \Spitzer\ systematics simultaneously with the exoplanet system parameters. The full model is described by:
\begin{equation}
    S(t)=\frac{\sum\limits^{n}_{i=1} w_i D_i(t)}{\sum\limits^{n}_{i=1} D_i(t)}+m \cdot t + M_{\mathrm{tr}}(\theta,t),\label{eq:1}
\end{equation}
where $w_i$ are individual time-independent pixel weights in the $n$ selected pixels in the region centered on the star, $D_i(t)$ is the observed flux (or counts) in the individual pixels of the selected region for each time step $t$, $m$ is the slope of a linear temporal ramp, and $M_{\mathrm{tr}}(\theta,t)$ is the transit model with model parameters $\theta$. The first part of this equation normalizes the individual pixel intensities so their sum at each time step is unity. In our analysis we used a $3\times3$ pixel region centered around the brightest pixel.

\begin{figure*}
    \centering
    \includegraphics[width=0.98\textwidth]{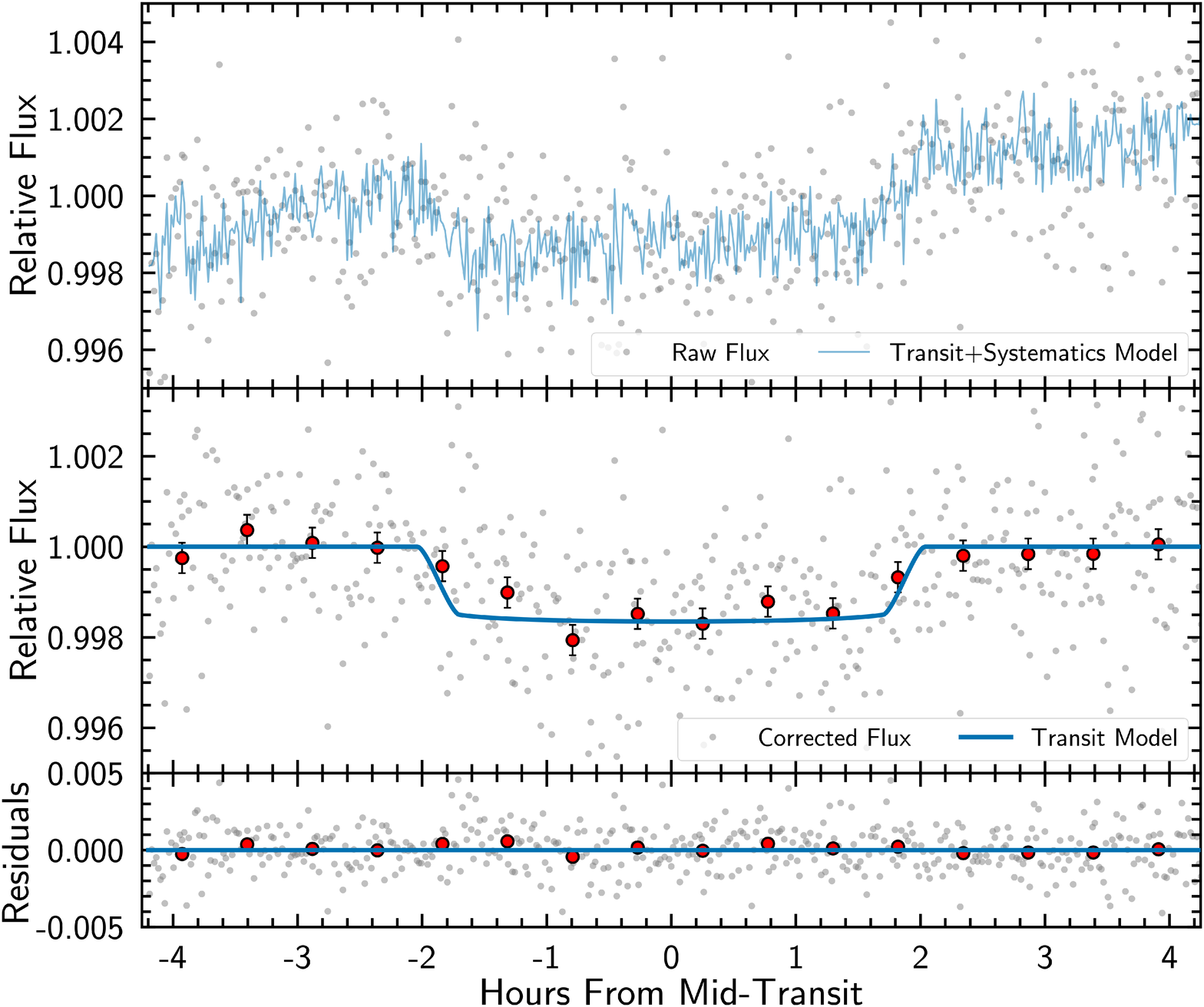}
    \caption{(Upper) Raw \Spitzer\ flux (gray points) with transit and systematics model for K2-138~g. (Middle) Systematics corrected flux (gray points) with transit model and $\sim$20 minute binned data (red) to highlight the drop in flux. (Lower) Residuals from the transit model fit.} \label{fig:spitzerlc}  
		\vspace{-0.5em}
\end{figure*}

\subsection{Transit Fitting}
\label{sec:transit}

Using \texttt{emcee} \citep{FM2013}, we simultaneously model the \kt\ and \Spitzer\ data, computing the posterior probability distributions for six transiting planets and the \Spitzer\ systematics. To model the transits we used \texttt{batman} \citep{Kreidberg2015}, which solves the analytic equations for an exoplanet transit as derived in \citet{Mandel2002}. We computed posterior probability distributions for the mid-transit times $T_0$, orbital periods $P$, the ratios of planet to star radii $R_p/R_{\star}$, the scaled semi-major axes $a/R_{\star}$, impact parameters $b$, two sets of quadratic limb darkening coefficients $q_1$ and $q_2$ (one set for \kt\ and another for \Spitzer), and the nine pixel weights $w_i$ and linear slope $m$ from Equation~\ref{eq:1}. We performed an autocorrelation analysis\footnote{\url{https://emcee.readthedocs.io/en/latest/tutorials/autocorr/}} to ensure chain convergence. Due to our large set of parameters, we used 500 walkers and 250,000 steps.

\begin{figure*}
    \centering
    \includegraphics[width=0.98\textwidth]{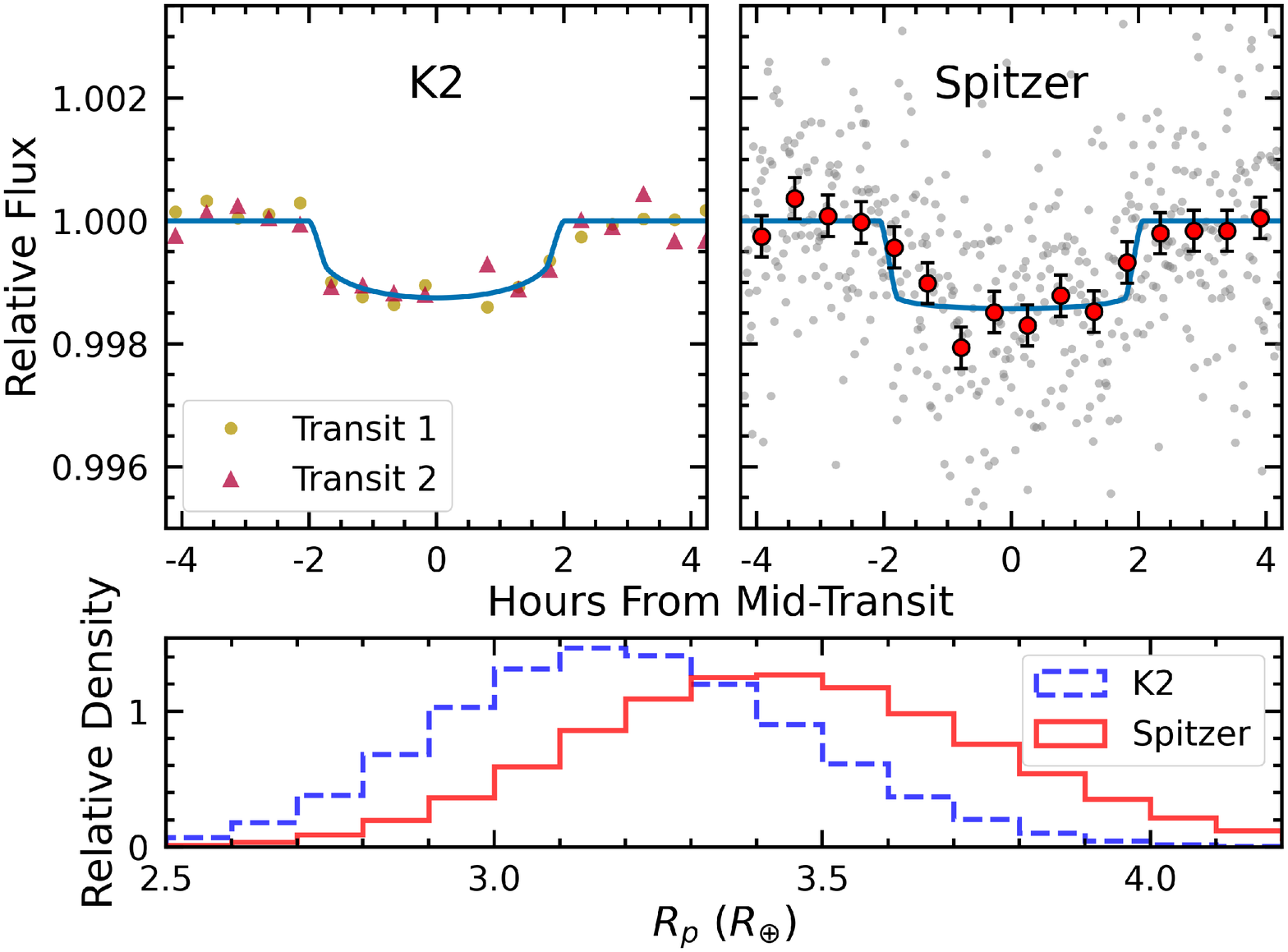}
    \caption{Transit comparison between \kt\ (upper left) and \Spitzer\ (upper right) on the same scale. The lower panel shows the computed radius posteriors in Earth radii for both \kt\ and \Spitzer. The \Spitzer\ radius is larger, but it is still consistent with the \kt\ radius within 1$\sigma$. It is also possible that systematics could bias the radius measurements. For example, having only two transits in the 30 minute cadence \kt\ data means that any outliers could skew the measured transit depth. Additional transits at these and other wavelengths will be necessary to constrain atmospheric properties of this planet.} \label{fig:g-rad}  
		\vspace{-0.5em}
\end{figure*}

The resultant \kt\ light curve fit using the median values of the posteriors is shown in Figure~\ref{fig:k2lc}, with the phase folded light curves shown in Figure~\ref{fig:k2lc-phase}. Figure~\ref{fig:spitzerlc} shows a clear transit event in the \Spitzer\ data for K2-138~g, confirming the existence of a sixth planet in the K2-138 system. We note that \citet{Christiansen2018} obtained high resolution AO imaging of the K2-138 system, ruling out nearby stellar companions that could contaminate or mimic a planet signal. Further, since K2-138 is a multi-planet system, it is more likely that additional transit-like signals come from another planet \citep[validation by multiplicity, e.g.,][]{Lissauer2014, Sinukoff2016}. Table~\ref{tab:planetpars} lists all the derived planet parameters for the \kt\ and \Spitzer\ data. We compare the \kt\ and \Spitzer\ light curves for K2-138~g in Figure~\ref{fig:g-rad}. The transit durations for the light curves are nearly identical, but the transit depth posterior distributions show a slightly larger radius in the \Spitzer\ data, although the difference is $< 1\sigma$.

\movetabledown=2.2in
\begin{rotatetable*}
\begin{deluxetable*}{ccccccccccccc}
\tabletypesize{\scriptsize} 
\tablecaption{K2-138 planet parameters.\label{tab:planetpars}}

\tablehead{\vspace{-10px}\\
\colhead{Planet} & \colhead{Period} & \colhead{$\mathrm{T}_0$} & \colhead{$\mathrm{T}_{14}$} & \colhead{$R_p$} & \colhead{$R_p$} & \colhead{$a$} & \colhead{$a$} & \colhead{$b$} & \colhead{$i$} & \colhead{$F$} & \colhead{$T_{\mathrm{eq}}^{\ \ \dagger}$} & \colhead{TSM} \\ \colhead{} & \colhead{d} & \colhead{BJD-2457700} & \colhead{hr} & \colhead{$R_{\star}$} & \colhead{$R_{\oplus}$} & \colhead{$R_{\star}$} & \colhead{AU} & \colhead{} & \colhead{$^{\circ}$} & \colhead{$F_{\oplus}$} & \colhead{K} & \colhead{}}
\startdata
b & $2.35321^{+0.00037}_{-0.00036}$ & $40.3713^{+0.0061}_{-0.0059}$ & $1.947^{+0.150}_{-0.168}$ & $0.01628^{+0.00116}_{-0.00111}$ & $1.49^{+0.15}_{-0.14}$ & $8.68^{+0.60}_{-0.58}$ & $0.0338^{+0.0034}_{-0.0031}$ & $0.35^{+0.21}_{-0.23}$ & $87.72^{+1.52}_{-1.54}$ & $424.51^{+126.30}_{-98.15}$ & $1157^{+78}_{-74}$ & $2.62^{+1.88}_{-0.98}$ \\
c & $3.56015^{+0.00022}_{-0.00023}$ & $40.3210^{+0.0029}_{-0.0028}$ & $2.339^{+0.089}_{-0.087}$ & $0.02463^{+0.00091}_{-0.00078}$ & $2.26^{+0.18}_{-0.17}$ & $11.39^{+0.53}_{-0.70}$ & $0.0442^{+0.0039}_{-0.0039}$ & $0.28^{+0.17}_{-0.18}$ & $88.59^{+0.92}_{-0.97}$ & $248.95^{+70.82}_{-55.02}$ & $1012^{+65}_{-61}$ & $25.91^{+10.33}_{-7.14}$ \\
d & $5.40484^{+0.00049}_{-0.00051}$ & $43.1569^{+0.0037}_{-0.0035}$ & $2.764^{+0.097}_{-0.086}$ & $0.02678^{+0.00093}_{-0.00081}$ & $2.46^{+0.19}_{-0.18}$ & $14.81^{+0.64}_{-0.86}$ & $0.0575^{+0.0050}_{-0.0049}$ & $0.23^{+0.17}_{-0.16}$ & $89.11^{+0.62}_{-0.75}$ & $147.12^{+41.30}_{-32.19}$ & $888^{+57}_{-53}$ & $23.25^{+8.74}_{-6.44}$ \\
e & $8.26147^{+0.00052}_{-0.00053}$ & $40.6451^{+0.0022}_{-0.0021}$ & $2.967^{+0.076}_{-0.070}$ & $0.03385^{+0.00110}_{-0.00094}$ & $3.11^{+0.24}_{-0.22}$ & $20.40^{+1.20}_{-1.57}$ & $0.0792^{+0.0076}_{-0.0077}$ & $0.38^{+0.15}_{-0.21}$ & $88.92^{+0.61}_{-0.55}$ & $77.79^{+23.51}_{-17.73}$ & $757^{+52}_{-47}$ & $23.27^{+8.27}_{-6.28}$ \\
f & $12.75758^{+0.00120}_{-0.00121}$ & $38.7018^{+0.0044}_{-0.0035}$ & $3.243^{+0.130}_{-0.111}$ & $0.02982^{+0.00130}_{-0.00122}$ & $2.73^{+0.23}_{-0.21}$ & $27.32^{+1.95}_{-2.11}$ & $0.1064^{+0.0108}_{-0.0105}$ & $0.48^{+0.12}_{-0.19}$ & $88.99^{+0.43}_{-0.35}$ & $43.02^{+13.24}_{-10.03}$ & $653^{+45}_{-42}$ & $23.45^{+32.85}_{-12.83}$ \\
$\mathrm{g}_{K2}$ & $41.96645^{+0.00603}_{-0.00665}$ & $73.8541^{+0.0044}_{-0.0044}$ & $4.035^{+0.156}_{-0.172}$ & $0.03472^{+0.00171}_{-0.00177}$ & $3.18^{+0.28}_{-0.26}$ & $58.59^{+4.84}_{-4.64}$ & $0.2287^{+0.0249}_{-0.0231}$ & $0.73^{+0.05}_{-0.07}$ & $89.29^{+0.11}_{-0.11}$ & $9.29^{+2.92}_{-2.23}$ & $445^{+31}_{-30}$ & $18.08^{+29.41}_{-10.83}$ \\
$\mathrm{g}_{Spitzer}^{\ \ \ \ \ \ \ \ \ \ddagger}$ & \nodata & $493.5155^{+0.0103}_{-0.0065}$ & $4.054^{+0.159}_{-0.175}$ & $0.03759^{+0.00214}_{-0.00230}$ & $3.44^{+0.32}_{-0.31}$ & \nodata & \nodata & \nodata & \nodata & \nodata & \nodata & \nodata \\
\enddata
\tablenotetext{\dagger}{Equilibrium temperatures were computed assuming a Bond albedo of 0.3.}
\tablenotetext{\ddagger}{Period, $a$, $b$, and $i$ were computed jointly with the \kt\ data.}
\end{deluxetable*}
\end{rotatetable*}

\section{Discussion}
\label{sec:discussion}

\subsection{Near Resonances and Gap Planets}
\label{sec:resonance}

\begin{figure}
    \centering
    \includegraphics[width=0.3095\textwidth]{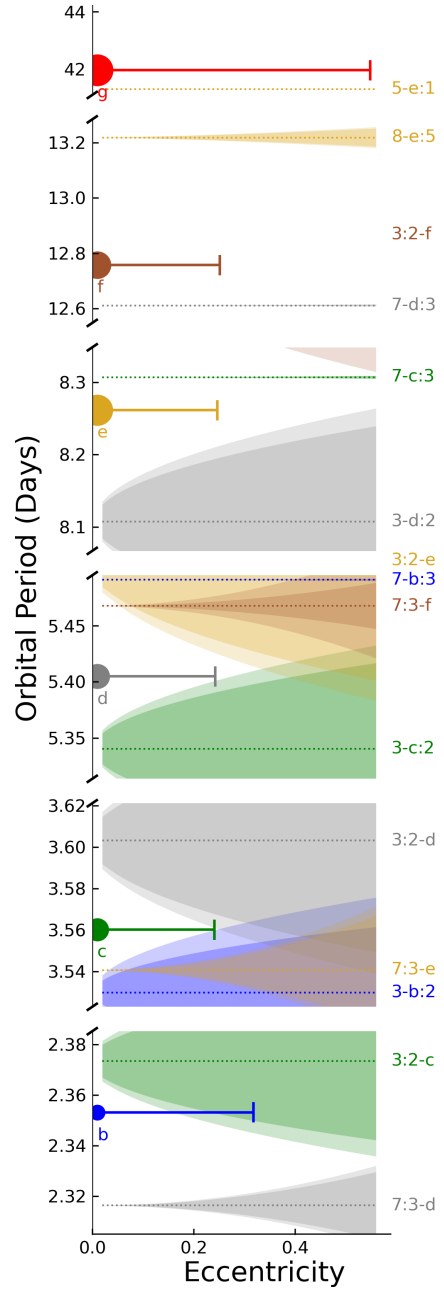}
    \caption{The location and analytically estimated widths of mean-motion resonances for the K2-138 system. Each planet is plotted in relative size to the other planets along the discontinuous y-axis indicating orbital period. Eccentricity is given along the x-axis, and extending from each planet is a line out to the eccentricity at which the planet would cross another planet's orbit. Horizontal dashed lines indicate the locations of interior (e.g., 3-b:2) and exterior (e.g., 3:2-c) resonances up to fourth order, color coded to match the label. The shaded regions surrounding each resonance line are the resonance widths corresponding to the lower (darker) and upper (lighter) planet mass limits. Planets b, c, d, and e are sufficiently near their mutual 3:2 resonances at low eccentricity for their dynamics to be affected, likely inducing TTVs (see Section~\ref{sec:massttv}).} \label{fig:res}  
		\vspace{-0.5em}
\end{figure}

The ratio of orbital periods between successive K2-138 planets are: c:b = 1.513, d:c = 1.518, e:d = 1.529, f:e = 1.544, g:f = 3.290. In order to determine how close to 3:2 resonance these planets are, we estimated the mean motion resonance widths using the program from \citet{Volk2020}\footnote{\url{https://github.com/katvolk/analytical-resonance-widths}}, which is based on the analytical derivations of resonance widths in the single-planet limit from \citet{Murray2000}. For this calculation, we used the masses of planets K2-138~b, c, d, and e from \citet{Lopez2019}, and estimated masses of K2-138~f and g ($6.72^{+8.04}_{-3.86}$ and $8.94^{+12.89}_{-5.91} M_{\oplus}$) from mass--radius relationships \citep{Ning2018}. The results of this calculation, out to fourth order mean motion resonances, are shown in Figure~\ref{fig:res}. Within the upper and lower planet mass limits, K2-138~b, c, d, and e are near (within a few half-widths) their mutual 3:2 resonances at low eccentricity, but the outer pair of planets are not near any low-order resonances.

The sizeable gap between K2-138~f and g leads to speculation that there could be additional non-transiting planets in the system. Indeed, \citet{Gilbert2020} suggest $\sim$20\% of high multiplicity planet systems host additional planets in the gaps between detected planets. Each consecutive planet pair of K2-138 has period ratios that slip further away from 3:2, and assuming the orbital period ratios continued at 1.544 (f:e), planets could be expected with orbital periods near 19.70, 30.42, and 46.98 days. However, if the K2-138 planets were all in perfect 3:2 resonance with planet b, there would orbits at 17.87, 26.80, and 40.21 days. Without additional data, we are unable to conclude whether or not K2-138~g would be near a 3:2 resonance with a planet in the gap.

Multi-planet systems have been found to be highly coplanar \citep[e.g.,][]{Fabrycky2014,Zhu2018,Gilbert2020}, however, the more distant a planet orbits, the closer to $90^{\circ}$ inclination it must be to be in a transiting geometry. Assuming orbital periods of 19.70 and 30.42 days, planets around K2-138 would need to be at inclinations above $88.9^{\circ}$ and $89.2^{\circ}$, respectively, for us to observe them in transit. Even within the Solar System, the planets are nearly coplanar, yet they still have mutual inclinations between $0\fdg33$ and $6.3^{\circ}$ \citep{Winn2015}.

\begin{figure*}
\gridline{\fig{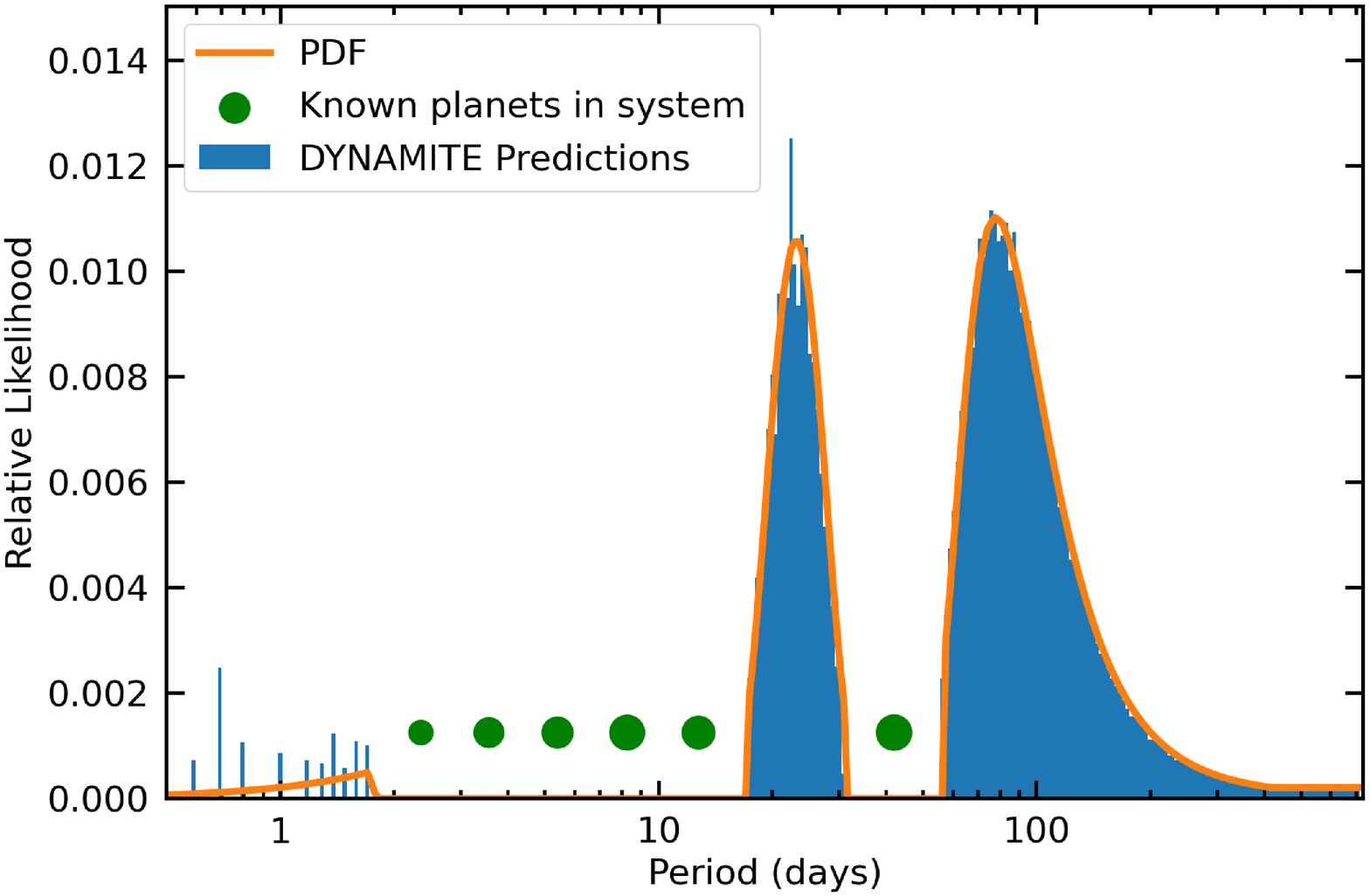}{0.48\textwidth}{(a)}
          \fig{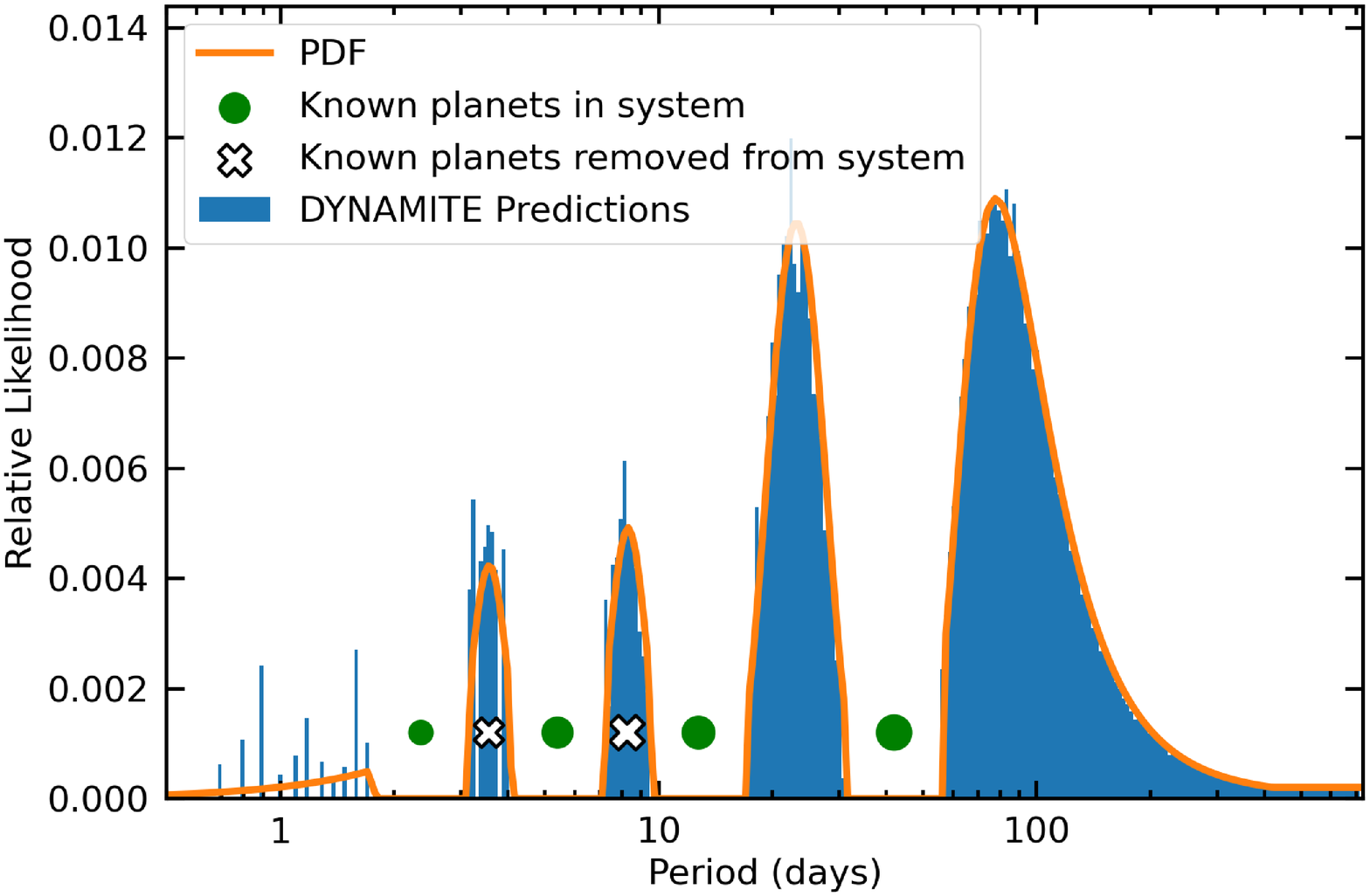}{0.48\textwidth}{(b)}}
\gridline{\fig{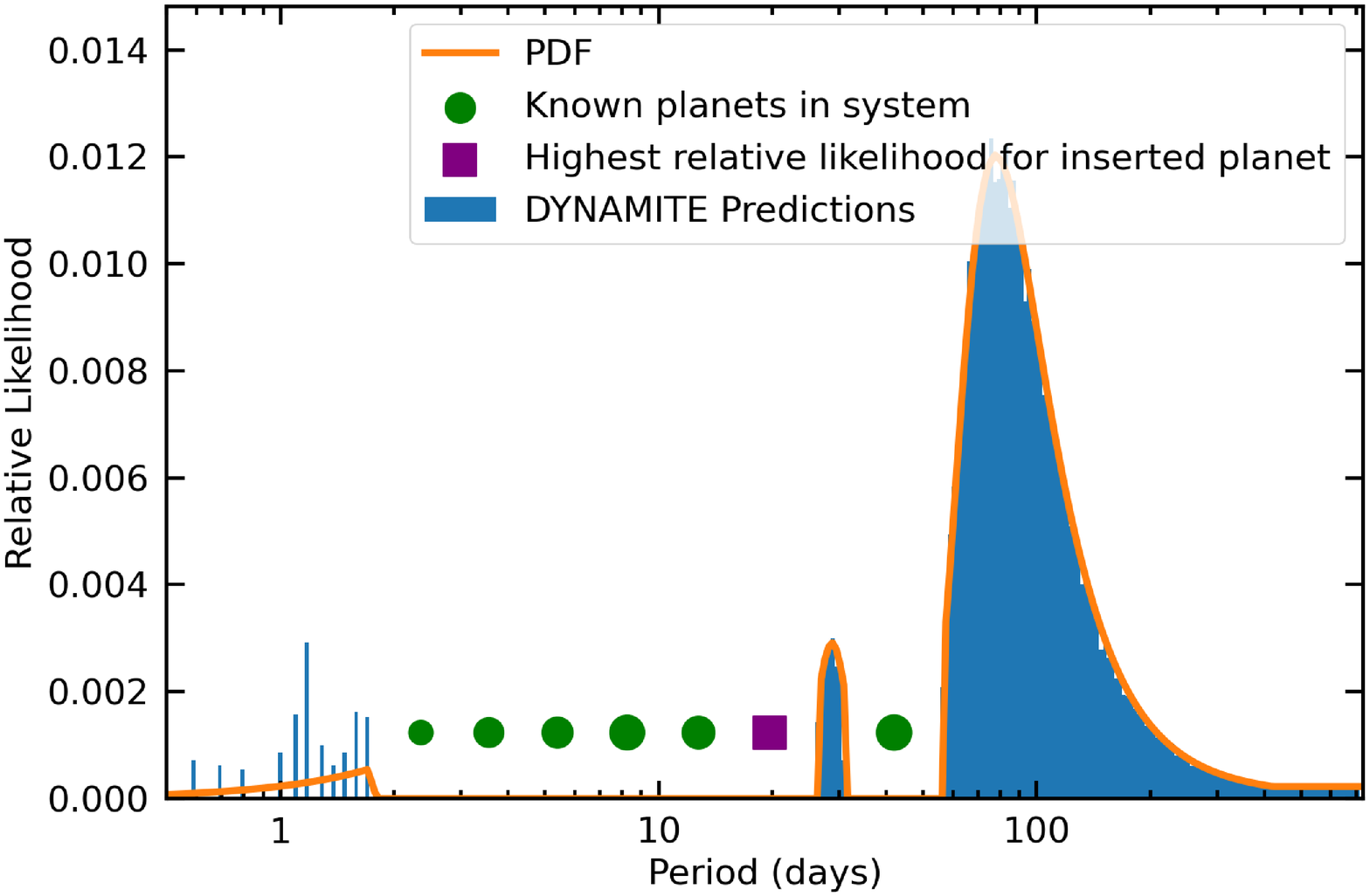}{0.48\textwidth}{(c)}
          \fig{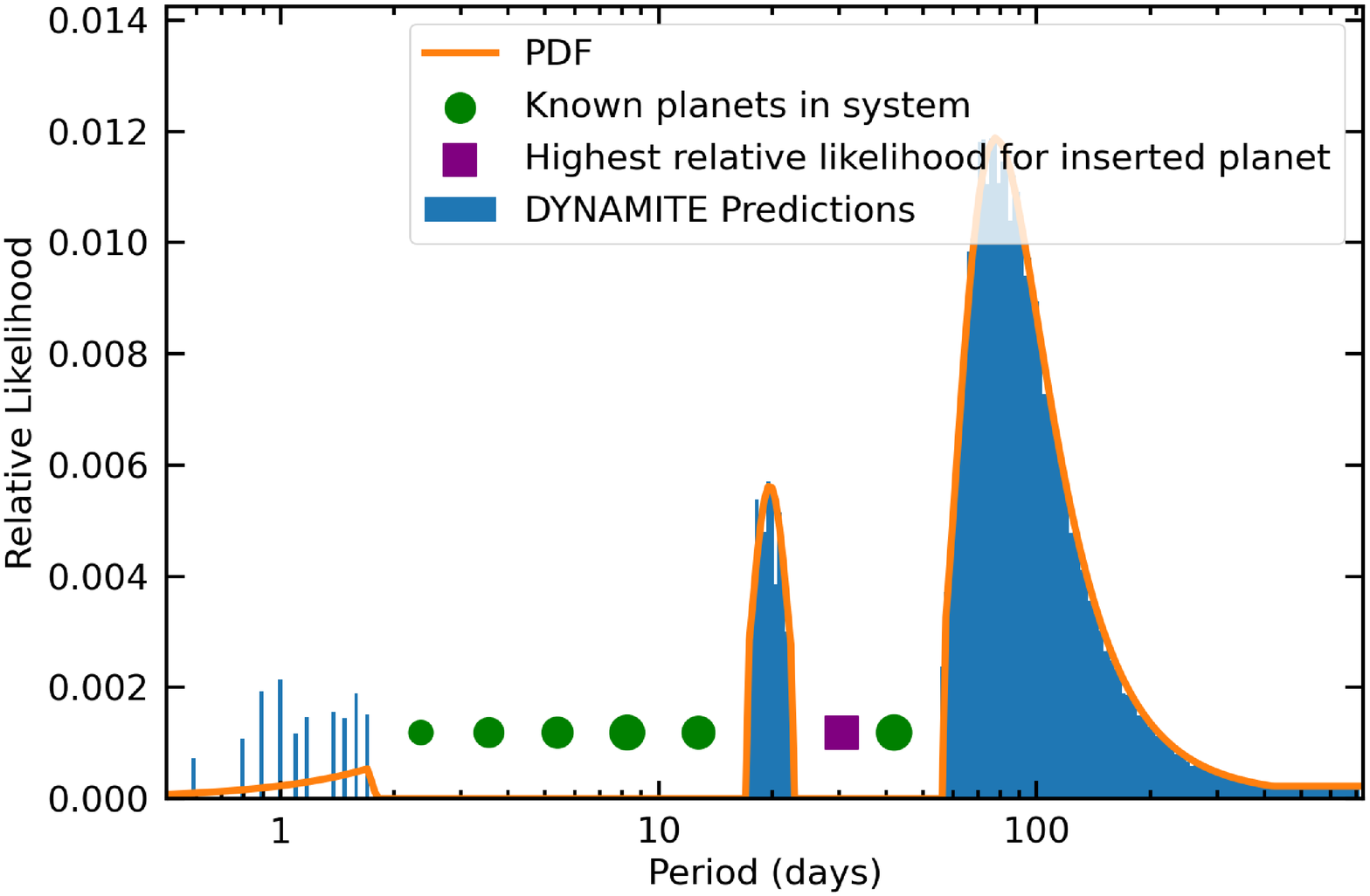}{0.48\textwidth}{(d)}}
\caption{\texttt{DYNAMITE} predictions of undetected planets. (a) Inputs of known K2-138 planets yields a prediction of a planet or planets with high relative likelihood in the gap between planets f and g, and additional planet(s) beyond g. (b) When planets c and e are removed, \texttt{DYNAMITE} predicts planets at their respective locations, indicating the predictive model yields the results we expect. (c) Injection of a planet at 19.70 days results in a planet prediction near 30 days, though with smaller relative likelihood than in the previous two scenarios. (d) Injection of a planet at 30.42 days yields a planet prediction near 20 days with a moderate relative likelihood.}\label{fig:prediction}
\end{figure*}

We further explore the possibility of planets within the gap between planets f and g using \texttt{DYNAMITE}\footnote{\url{https://github.com/JeremyDietrich/dynamite}}, which uses population statistics to predict previously undetected planets \citep{Dietrich2020}. This model takes inputs of stellar parameters (radius, mass, temperature) and known planet parameters (inclination, radius, period), and yields probability distributions where the population models predict a planet or planets might exist. We considered four different scenarios as inputs to \texttt{DYNAMITE}, which are shown in Figure~\ref{fig:prediction}: (a) all currently known/detected K2-138 planets, (b) removal of K2-138~c and e, (c) all K2-138 planets with a planet injected at 19.70 days, and (d) all K2-138 planets with a planet injected at 30.42 days. In scenario (a), \texttt{DYNAMITE} predicted a planet or planets to be within the gap between K2-138~f and g, and beyond K2-138~g. The model accurately predicted the locations of K2-138~c and e in scenario (b). From our planet injection tests in scenarios (c) and (d), the models predicted a planet near 30 and 20 days, respectively.

\subsection{Masses and TTVs}
\label{sec:massttv}

Due to its distance from first order resonance, TTV measurements for K2-138~g would be difficult. Using \texttt{TTVFaster} \citep{Agol2016} we computed TTV amplitudes of $2.23^{+0.38}_{-0.43}$, $4.70^{+1.13}_{-1.11}$, $8.57^{+1.40}_{-1.46}$, $6.92^{+3.84}_{-2.18}$, $7.17^{+1.23}_{-1.13}$, and $0.09^{+0.12}_{-0.05}$ minutes for K2-138~b, c, d, e, f, and g, respectively. Our inputs to \texttt{TTVFaster} were the masses of planets K2-138~b, c, d, and e from \citet{Lopez2019} and the aforementioned estimated masses of K2-138~f and g. We also assumed zero eccentricity. Our average six minute ($1\sigma$) \kt\ timing precision was insufficient to measure TTVs for this system, however, higher cadence (one minute) observations with CHEOPS should improve the timing precision enough to allow detection of TTVs of the inner five planets.

In measuring the masses of the inner four K2-138 planets, \citet{Lopez2019} did not identify additional planets, though additional signals might have been absorbed by their Gaussian process to fit out stellar activity at the 5.6\,m\,s$^{-1}$ level. The mass measurement of K2-138~f was hindered by its orbital period of 12.8 days, near half the 24.7 day stellar rotation period. The stellar rotation period might also hinder detection of a planet in orbit near the next 3:2 resonance beyond planet f around 20 days. Future planet searches and mass measurements for this system would likely benefit from simultaneous photometric and RV observations, as \citet{Kosiarek2020} suggest this could enhance the precision of RV measurements. \citet{Lopez2019} were unable to reliably measure a mass of K2-138~g either, but assuming a mass of $8.94^{+12.89}_{-5.91}\,M_{\oplus}$, we predict an RV semi-amplitude of $1.79^{+2.56}_{-1.18}$\,m\,s$^{-1}$. If there were planets between f and g of similar masses to the other planets in the system, we would expect them to have RV semi-amplitudes between 1.5 and 2.5\,m\,s$^{-1}$, which would make them similarly difficult to detect due to stellar activity levels.

\begin{figure}
    \centering
    \includegraphics[width=0.45\textwidth]{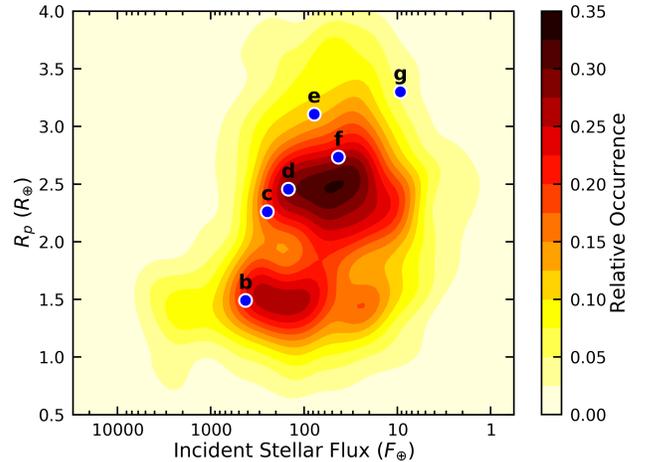}
    \caption{K2-138 planet radii vs.\ incident stellar flux. The contours represent the population of 816 confirmed and candidate \kt\ planets from \citet{Hardegree-Ullman2020}. The high incident stellar flux on K2-138~b likely stripped away its atmosphere, placing it below the planet radius valley, whereas the other planets were able to maintain their atmospheres.} \label{fig:insfl}  
		\vspace{-0.5em}
\end{figure}

We note that the outer five planets of K2-138 are all sub-Neptunes similar in size, and planet b is likely a rocky super-Earth with a density of $5.01^{+2.73}_{-2.00}$\,g\,cm$^{-3}$. Common sizing of multi-planet systems has previously been found for \Kepler\ systems \citep[e.g.][]{Millholland2017,Wang2017,Weiss2018,Gilbert2020}. From a planet formation standpoint, \citet{Adams2020} found that energy optimization occurs when planets are nearly equal in mass for low-mass (super-Earth/sub-Neptune) planet systems, which is consistent with what we see with K2-138. Though, we note that the outer planets of K2-138 have larger radii than the inner planets, a trend consistent with the findings of \citet{Ciardi2013}, \citet{Millholland2017}, \citet{Kipping2018}, and \citet{Weiss2018}, possibly the result of enhanced photoevaporation closer to the star. We plot the planet radii with respect to incident stellar flux for the K2-138 planets, compared to the population of \kt\ planets (shown as density contours) from \citet{Hardegree-Ullman2020} in Figure~\ref{fig:insfl}. K2-138~b has incident flux ($F_{\oplus}$) over 400 times higher than Earth and is the only planet in the system below the planet radius valley. The other planets in the system receive less than 250 $F_{\oplus}$, apparently low enough to retain an atmosphere.

From Figure~\ref{fig:insfl}, it appears that many of the K2-138 planets are inflated relative to their counterparts with similar incident stellar flux. If the system was relatively young, we would expect the planets to still be undergoing mass loss. \citet{Lopez2019} computed an age of $2.3^{+0.44}_{-0.36}$ Gyr for K2-138 based on chromospheric emission, and $2.8^{+3.8}_{-1.7}$ Gyr from their joint radial velocity, light curve, and spectral energy distribution analysis. Similarly, we input photometry, stellar parameters, and a rotation period of 24.7 days into the isochrone fitting with gyrochronology package \texttt{stardate}\footnote{\url{https://github.com/RuthAngus/stardate}} and compute an age of $2.8\pm0.3$ Gyr, consistent with \citet{Lopez2019}. However, a visual assessment of the raw flux in Figure~\ref{fig:k2lc} and a Lomb-Scargle periodogram yields a significant peak corresponding to a period of $\sim$12.5 days. This rotation period corresponds to a younger age closer to $\sim$0.9 Gyr. We note that even at this younger age, it is unlikely the planets are still undergoing significant mass loss since this process occurs within the first few hundred Myrs \citep{Lopez2012}.

\begin{figure*}[ht!]
    \centering
    \includegraphics[width=\textwidth]{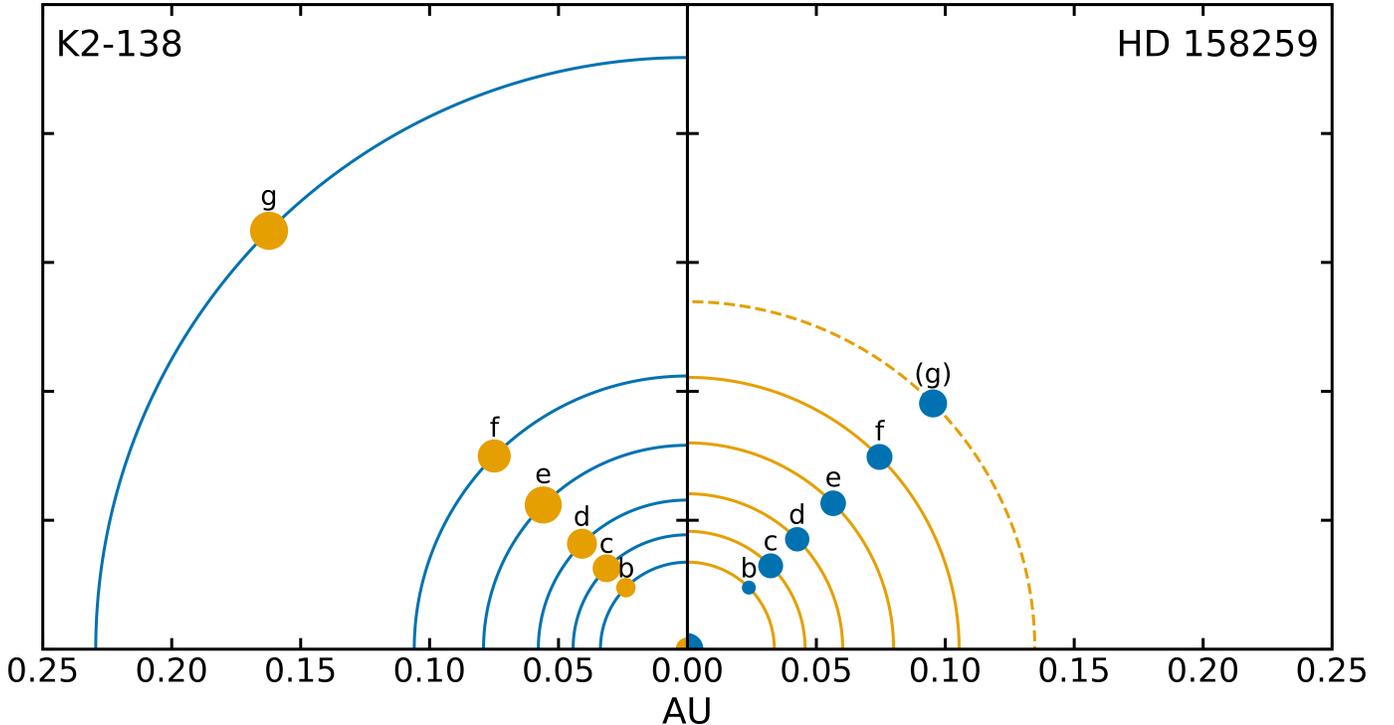}
    \caption{The orbits and planets of K2-138 and HD~158259, highlighting the similarities of the two systems. The orbits and the star sizes are to scale, but the planets are enlarged by 50$\times$ to show detail. Orbital distances of 0.034, 0.046, 0.060, 0.080, 0.105, and 0.135 AU for HD~158259 b, c, d, e, f, and tentative planet (g), respectively, were computed from Kepler's third law using stellar mass and planet orbital periods from \citet{Hara2020}.} \label{fig:k2hd}
		\vspace{-0.25em}
\end{figure*}

Another possibility for these relatively large planets is tidally induced radius inflation \citep{Millholland2019b}. The \Kepler\ mission unveiled a statistical overabundance of planet pairs just outside of first-order mean motion resonances, specifically 2:1 and 3:2 \citep{Lissauer2011,Millholland2019a}. As noted in Section~\ref{sec:resonance}, most of the planet pairs of K2-138 fall just outside a 3:2 resonance. Tidal forces from the host star can push planets into near-resonant configurations, but host-star tides alone cannot explain how all the energy from this process is dissipated to keep planets in this configuration. \citet{Millholland2019a} showed that obliquity tides may be the source of energy dissipation that helps sculpt these near-resonant systems. Consequently, these tidal forces heat the planet interiors, leading to atmospheric inflation \citep{Millholland2019b}. We posit that K2-138 is a strong candidate for planet radius inflation due to obliquity tides.

\subsection{Comparison to Other Multi-planet Systems}
\label{sec:comparison}

To date, there have only been nine other exoplanet systems with six or more confirmed planets\footnote{\href{https://exoplanetarchive.ipac.caltech.edu/cgi-bin/TblView/nph-tblView?app=ExoTbls&config=PS}{https://exoplanetarchive.ipac.caltech.edu/cgi-bin/TblView/}
\\
\href{https://exoplanetarchive.ipac.caltech.edu/cgi-bin/TblView/nph-tblView?app=ExoTbls&config=PS}{nph-tblView?app=ExoTbls\&config=PS}, as of February 2021.}, including radial velocity discovered systems HD~10180 (6 planets), HD~219134 (6 planets), and HD~34445 (6 planets), and transiting systems Kepler-11 (6 planets), Kepler-20 (6 planets), Kepler-80 (6 planets), Kepler-90/KOI-351 (8 planets), TRAPPIST-1 (7 planets), and TOI-178 (6 planets). Perhaps most similar to K2-138, however, is the HD~158259 system, with five confirmed planets and a sixth candidate outer planet \citep{Hara2020}, all near 3:2 orbital mean motion resonances. Four of the confirmed planets were detected in radial velocity data with the SOPHIE spectrograph, and the innermost planet was found to be transiting in \textit{TESS} data. The outermost candidate planet orbits every 17.4 days, close to the stellar rotation period, complicating confirmation of this planet. The five innermost planets of K2-138 and HD~158259 are each located at nearly identical distances to their host stars, as shown in Figure~\ref{fig:k2hd}. We estimated HD~158259 planet radii for the five non-transiting planets using the planet masses from \citet{Hara2020} and the mass-radius relationships of \citet{Chen2017}. These non-transiting planets are also all similar-sized sub-Neptunes, with estimated radii larger than $2\,R_{\oplus}$--again consistent with the aforementioned common sizing of multi-planet systems. Each respective planet in HD~158259 is slightly smaller than its counterpart K2-138 planet, which could be the result of HD~158259 being a larger host star ($1.08\pm0.10\,M_{\odot}$) that is more efficient at stripping away planetary atmospheres by intense irradiation \citep{Ehrenreich2015}. We note, however, that there are significant uncertainties in planetary mass-radius relationships. Without transit data, it is difficult to test whether or not this system undergoes tidal radius inflation as mentioned in Section~\ref{sec:massttv}.

\begin{figure*}[ht!]
    \centering
    \includegraphics[width=\textwidth]{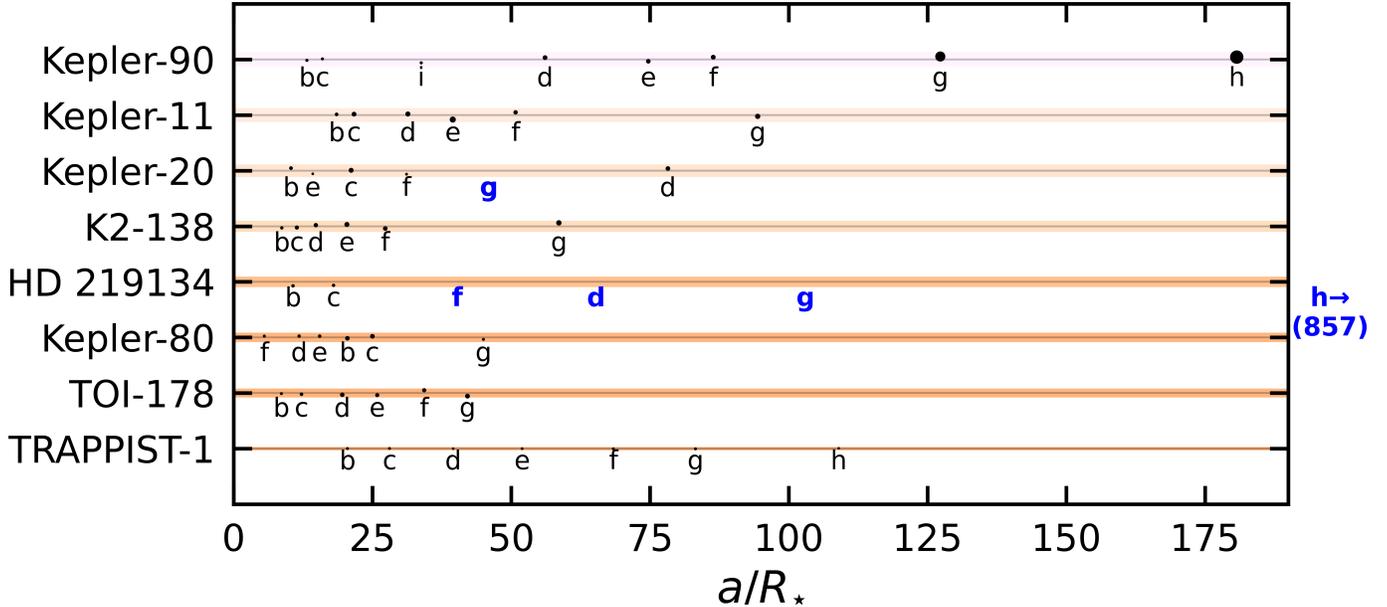}
    \caption{Orbital spacing of systems with six or more planets, and at least one transiting planet. Systems are arranged from largest (top) to smallest (bottom) stellar host, and the regions are colored according to the host temperatures \citep{Harre2021}. The width of the colored regions are scaled to the stellar radii. Planets are to scale with the stellar radii but enlarged by $10\times$ for clarity, and are placed at their respective transiting inclination angles (randomly distributed above and below the stellar mid-point; gray line). Non-transiting planet locations are labeled in blue. For clarity, we did not plot HD~219134\,h on this scale, but note that it is located at $a/R_{\star}=857$. In addition to K2-138, the systems Kepler-11, Kepler-20, HD~219134, and Kepler-80 have notable gaps between their outermost planets.} \label{fig:spacing}
		\vspace{-0.25em}
\end{figure*}

We qualitatively compared the orbital spacing ($a/R_{\star}$) of these high-multiplicity systems with transiting planets (Figure~\ref{fig:spacing}). In addition to the K2-138 system, there is a sizeable gap between the outermost detected transiting planets of the Kepler-11, Kepler-20, and Kepler-80 systems. HD~219134 has two transiting planets and four non-transiting planets detected via RV measurements, again with a large gap between the two outermost planets. This large outermost planet gap is also present in the RV system HD~34445. Notably, a non-transiting planet was identified in the gap between outer planets Kepler-20 f and d with RV data \citep{Buchhave2016}. As noted in Section~\ref{sec:resonance}, planets orbiting further out must be closer to $i=90^{\circ}$ to be in a transiting geometry, but RV and TTV data may uncover unseen planets. We encourage further investigations of this outer planet gap feature in high-multiplicity planet systems in order to disambiguate whether it is caused by observational biases or planet formation processes.

\subsection{JWST, ARIEL, and Future Prospects}
\label{sec:JWST}
    
We computed the transmission spectroscopy metric (TSM) for the K2-138 planets as defined by Equations 1, 2, and 3 of \citet{Kempton2018}. The TSM is the expected signal-to-noise for a 10 hour observing program with JWST/NIRISS. For planets b, c, d, and e, we used the planet masses measured by \citet{Lopez2019}, and for planets f and g, we used our estimated masses. The equilibrium temperature was calculated assuming zero albedo and full day-night heat redistribution. The resultant TSM values are listed in Table~\ref{tab:planetpars}, and are $\sim$20 for the outer planets, falling to 2.62 for the innermost planet b. These values are well below the recommended threshold of TSM $>$ 90 for high quality atmospheric characterization of sub-Neptune sized planets. For now, the K2-138 planets are unlikely to be selected as high-priority targets for JWST observations.

The European Space Agency Atmospheric Remote-sensing Infrared Exoplanet Large-survey (ARIEL) space mission aims to gather transmission spectra of 1000 exoplanets during its four year mission (expected to launch in 2028) in order to study their composition, formation, and evolution. \citet{Edwards2019} compiled a list of potential targets for ARIEL, taking into account currently known stellar and planet parameters. K2-138~falls very near the average star system considered for this target list. The inner five planets of K2-138 also fall within the range of planets considered for the potential target list, but very few planets with orbital periods beyond $\sim$20 days will likely be considered, all but ruling out observations of K2-138~g. However, since K2-138~contains five similarly-sized sub-Neptunes with a $\sim$500 K range of equilibrium temperatures from warm to temperate, these planets might provide a unique test bed for comparative sub-Neptune atmosphere studies.

We have confirmed the existence of K2-138~g, solidifying K2-138 as the largest \kt\ multi-planet system. K2-138~g breaks the continuous near 3:2 mean motion resonance of the inner five planets, but the sizeable gap between K2-138 f and g hints at the possibility there could be additional non-transiting planets in this system. We encourage future observations of this potential key benchmark system to (1) constrain TTVs of the inner planets, (2) enable more precise masses and potential discovery of additional planets with simultaneous photometric and RV measurements, and (3) facilitate comparative atmospheric studies of warm to temperate sub-Neptune planets.

\section{Acknowledgements}
\label{sec:ack}
We would like to thank Matt Russo and SYSTEM Sounds for their creative sonification of the K2-138 system: \url{http://www.system-sounds.com/k2-138/}. KKHU would like to thank Jon Zink for helpful discussions regarding transit fitting.

This work is based in part on observations made with the {\it Spitzer Space Telescope}, which was operated by the Jet Propulsion Laboratory, California Institute of Technology under a contract with NASA.

This paper includes data collected by the \kt\ mission. Funding for the \kt\ mission is provided by the NASA Science Mission directorate.

This research has made use of the NASA Exoplanet Archive, which is operated by the California Institute of Technology, under contract with the National Aeronautics and Space Administration under the Exoplanet Exploration Program.

This research made use of Astropy,\footnote{http://www.astropy.org} a community-developed core Python package for Astronomy \citep{astropy2013, astropy2018}.

This research made use of \texttt{exoplanet} \citep{exoplanet:exoplanet} and its dependencies \citep{exoplanet:agol19, exoplanet:astropy13, exoplanet:astropy18, exoplanet:exoplanet, exoplanet:foremanmackey17, exoplanet:foremanmackey18, exoplanet:kipping13, exoplanet:luger18, exoplanet:pymc3, exoplanet:theano, exoplanet:vaneylen19}.

KV acknowledges funding from NASA (grant 80NSSC18K0397).

\facilities{Spitzer, Kepler, Exoplanet Archive}

\software{\texttt{Astropy} \citep{astropy2013,astropy2018}, \texttt{batman} \citep{Kreidberg2015}, \texttt{dustmaps} \citep{Green2018}, \texttt{DYNAMITE} \citep{Dietrich2020}, \texttt{emcee} \citep{FM2013}, \texttt{exoplanet} \citep{exoplanet:exoplanet}, \texttt{isoclassify} \citep{Huber2017}, \texttt{NumPy} \citep{numpy2020}, \texttt{photutils} \citep{Bradley2019}, \texttt{SciPy} \citep{SciPy-NMeth2020}, \texttt{TTVFaster} \citep{Agol2016}}

\bibliography{K2-138}\setlength{\itemsep}{-2mm}

\end{document}